\documentclass[preprint,aps,prc,superscriptaddress,showpacs,floatfix]{revtex4}
\usepackage{graphicx}
\usepackage{dcolumn}
\usepackage{bm}

\begin{document}

\preprint{}
\title{Single and double $\pi ^{-}/\pi ^{+}$ ratios in heavy-ion reactions
as probes of the high-density behavior of the nuclear symmetry energy}
\author{Gao-Chan Yong}
\affiliation{Institute of Modern Physics, Chinese Academy of Science, Lanzhou 730000,
China}
\affiliation{Graduate School, Chinese Academy of Science, Beijing 100039, P.R. China}
\author{Bao-An Li}
\affiliation{Department of Chemistry and Physics, P.O. Box 419, Arkansas State
University, State University, Arkansas 72467-0419, USA }
\author{Lie-Wen Chen}
\affiliation{Institute of Theoretical Physics, Shanghai Jiao Tong University, Shanghai
200240, China}
\affiliation{Center of Theoretical Nuclear Physics, National Laboratory of Heavy Ion
Accelerator, Lanzhou 730000, China }
\author{Wei Zuo}
\affiliation{Institute of Modern Physics, Chinese Academy of Science, Lanzhou 730000,
China}

\begin{abstract}
Based on an isospin- and momentum-dependent hadronic transport model IBUU04,
effects of the nuclear symmetry energy on the single and double $\pi
^{-}/\pi ^{+}$ ratios in central reactions of $^{132}$Sn+$^{124}$Sn and $%
^{112}$Sn+$^{112}$Sn at a beam energy of $400$ MeV/nucleon are studied. It
is found that around the Coulomb peak of the single $\pi ^{-}/\pi ^{+}$
ratio the double $\pi ^{-}/\pi ^{+}$ ratio taken from the two isotopic
reactions retains about the same sensitivity to the density dependence of
nuclear symmetry energy. Because the double $\pi ^{-}/\pi ^{+}$ ratio can
reduce significantly the systematic errors, it is thus a more effective
probe for the high-density behavior of the nuclear symmetry energy.
\end{abstract}

\maketitle

\section{Introduction}
After about three decades of intensive efforts in both nuclear experiments
and theories, the equation of state (EOS) of isospin symmetric nuclear
matter is now relatively well determined mainly by studying collective flows
in heavy-ion collisions \cite{pd02} and nuclear giant monopole resonances
\cite{youngblood99}. The major remaining uncertainty about the EOS of
symmetric nuclear matter is due to our poor knowledge about the density
dependence of the nuclear symmetry energy \cite{pd02,pie04,colo04}.
Therefore, the new challenge is to determine the EOS of isospin asymmetric
nuclear matter, especially the density dependence of the nuclear symmetry
energy. Besides the great interests in nuclear physics, the EOS of
asymmetric nuclear matter is also crucial in many astrophysical processes,
especially in connection with the structure of neutron stars and the
dynamical evolution of proto-neutron stars \cite{mk94}. Fortunately,
heavy-ion reactions, especially those induced by radioactive beams, provide
a unique opportunity to constrain the EOS of asymmetric nuclear matter \cite%
{ireview,ibook,baran05}. In fact, considerable progress has been made
recently in determining the density dependence of the nuclear symmetry
energy around the normal nuclear matter density from studying the isospin
diffusion in heavy-ion reactions at intermediate energies \cite%
{mbt,chen04,li05}. However, much more work is still needed to probe the
high-density behavior of the nuclear symmetry energy.

A crucial task is to find experimental observables that are sensitive to the
density dependence of the nuclear symmetry energy. A number of such
observables have been already identified in heavy-ion collisions induced by
neutron-rich nuclei, such as the free neutron/proton ratio \cite{ba97a}, the
isospin fractionation \cite{serot,ba97b,vb98,hs00,wp01,vb02}, the
neutron-proton transverse differential flow \cite{ba00a,vg03,ls99}, the
neutron-proton correlation function \cite{lw03a}, $t$/$^{3}$He \cite%
{chen03a,zhang05}, the isospin diffusion \cite{lw04a,ls03}, the
proton differential elliptic flow \cite{ba01a} and the $\pi
^{-}/\pi ^{+}$ ratio \cite{ba02a,gai04,qli05a,qli05b}. Generally,
the long range Coulomb force on charged particles plays an
important role in the above observables. It is thus important to
distinguish the effects due to the symmetry potentials from those
due to the Coulomb potentials. It is also useful to understand the
interplay between these two kinds of potentials. Moreover, to
extract accurately information about the symmetry energy one has
to reduce as much as possible the systematic errors involved in
the observables used in experiments. For this purpose one normally
studies the ratios or relative values of experimental observables
from two reaction systems using different isotopes of the same
element. The first theoretical investigation of such an
observable, the double neutron/proton ratio of pre-equilibrium
nucleons, has been made recently in Ref. \cite{li05b}. Since the
single $\pi ^{-}/\pi ^{+}$ ratio in heavy-ion collisions induced
by neutron-rich nuclei has been shown to be a useful probe of the
high-density behavior of the nuclear symmetry energy
\cite{ba02a,qli05a,qli05b}, we study here the effects of the
symmetry
energy on the double $\pi ^{-}/\pi ^{+}$ ratio from the reactions of $^{132}$%
Sn+$^{124}$Sn and $^{112}$Sn+$^{112}$Sn, i.e., the ratio of $\pi
^{-}/\pi ^{+}$ from $^{132}$Sn+$^{124}$Sn over that from
$^{112}$Sn+$^{112}$Sn, using the IBUU04 model. It is well known
that the single $\pi ^{-}/\pi ^{+}$ ratio has a Coulomb peak at
certain pion kinetic energy depending on the system and the impact
parameter of the reaction \cite{ben79,bert80,lib79,gyu81}. It is
thus especially interesting to examine the sensitivity of the
double $\pi ^{-}/\pi ^{+}$ ratio to the symmetry energy around the
Coulomb peak. We find that around the Coulomb peak the double $\pi
^{-}/\pi ^{+}$ ratio has about the same sensitivity to the
symmetry energy as the single $\pi ^{-}/\pi ^{+}$ ratio while
having the advantage of reduced systematic errors.

\section{A BRIEF INTRODUCTION TO THE IBUU04 TRANSPORT MODEL}
In the IBUU04 model, besides nucleons, $\Delta $ and $N^{\ast }$
resonances as well as pions and their isospin-dependent dynamics
are included. The initial neutron and proton density distributions
of the projectile and target are obtained by using the
relativistic mean field theory. The experimental free-space
nucleon-nucleon (NN) scattering cross sections and the in-medium
NN cross sections can be used optionally. In the present work, we
use the isospin-dependent in-medium NN elastic cross sections from
the scaling model according to nucleon effective masses
\cite{li05}. For the inelastic cross sections we use the
experimental data from free space NN collisions since the
in-medium inelastic NN cross sections are still very much
controversial. The total and differential cross sections for all
other particles are taken either from experimental data or
obtained by using the detailed balance formula. The isospin
dependent phase-space distribution functions of the particles
involved are solved by using the test-particle method numerically.
The isospin-dependence of Pauli blockings for fermions
is also considered. More details can be found in Refs. \cite%
{ba97a,ba04a,das03,ba04,li05}. The momentum-dependent single nucleon
potential (MDI) adopted here is \cite{das03}%
\begin{eqnarray}
U(\rho ,\delta ,\mathbf{p},\tau ) &=&A_{u}(x)\frac{\rho _{\tau ^{\prime }}}{%
\rho _{0}}+A_{l}(x)\frac{\rho _{\tau }}{\rho _{0}}  \nonumber \\
&&+B(\frac{\rho }{\rho _{0}})^{\sigma }(1-x\delta ^{2})-8x\tau \frac{B}{%
\sigma +1}\frac{\rho ^{\sigma -1}}{\rho _{0}^{\sigma }}\delta \rho _{\tau
^{\prime }}  \nonumber \\
&&+\frac{2C_{\tau ,\tau }}{\rho _{0}}\int d^{3}\mathbf{p}^{\prime }\frac{%
f_{\tau }(\mathbf{r},\mathbf{p}^{\prime })}{1+(\mathbf{p}-\mathbf{p}^{\prime
})^{2}/\Lambda ^{2}}  \nonumber \\
&&+\frac{2C_{\tau ,\tau ^{\prime }}}{\rho _{0}}\int d^{3}\mathbf{p}^{\prime }%
\frac{f_{\tau ^{\prime }}(\mathbf{r},\mathbf{p}^{\prime })}{1+(\mathbf{p}-%
\mathbf{p}^{\prime })^{2}/\Lambda ^{2}}.  \label{potential}
\end{eqnarray}%
In the above equation, $\delta =(\rho _{n}-\rho _{p})/(\rho _{n}+\rho _{p})$
is the isospin asymmetry parameter, $\rho =\rho _{n}+\rho _{p}$ is the
baryon density and $\rho _{n},\rho _{p}$ are the neutron and proton
densities, respectively. $\tau =1/2(-1/2)$ for neutron (proton) and $\tau
\neq \tau ^{\prime }$, $\sigma =4/3$, $f_{\tau }(\mathbf{r},\mathbf{p})$ is
the phase-space distribution function at coordinate $\mathbf{r}$ and
momentum $\mathbf{p}$. The parameters $A_{u}(x),A_{l}(x),B,C_{\tau ,\tau }$,
$C_{\tau ,\tau ^{\prime }}$ and $\Lambda $ were set by reproducing the
momentum-dependent potential $U(\rho ,\delta ,\mathbf{p},\tau )$ predicted
by the Gogny Hartree-Fock and/or the Brueckner-Hartree-Fock calculations,
the saturation properties of symmetric nuclear matter and the symmetry
energy of about $32$ MeV at normal nuclear matter density $\rho _{0}=0.16$ fm%
$^{-3}$. The incompressibility of symmetric nuclear matter at normal density
is set to be $211$ MeV. The parameters $A_{u}(x)$ and $A_{l}(x)$ depend on
the x parameter according to
\begin{equation}
A_{u}(x)=-95.98-\frac{2B}{\sigma +1}x,\text{ }A_{l}(x)=-120.57+\frac{2B}{%
\sigma +1}x,
\end{equation}%
where $B=106.35$ MeV. $\Lambda =p_{F}^{0}$ is the nucleon Fermi momentum in
symmetric nuclear matter, $C_{\tau ,\tau ^{\prime }}=-103.4$ MeV and $%
C_{\tau ,\tau }=-11.7$ MeV. The $C_{\tau ,\tau ^{\prime }}$ and $C_{\tau
,\tau }$ terms are the momentum-dependent interactions of a nucleon with
unlike and like nucleons in the surrounding nuclear matter. The parameter
\emph{x} is introduced to mimic various density-dependent symmetry energies $%
E_{\text{sym}}(\rho )$ predicted by microscopic and phenomenological
many-body approaches \cite{diep03}. The isoscalar part $(U_{n}+U_{p})/2$
\cite{ba04a,ba04} of the single nucleon potential was shown to be in good
agreement with that of the variational many-body calculations \cite{rbw88}
and the results of the BHF approach including three-body forces \cite{zuo05}%
. The isovector part $(U_{n}-U_{p})/2\delta $ \cite{ba04a} is
consistent with the experimental Lane potential \cite{am62}.
\begin{figure}[th]
\begin{center}
\includegraphics[width=0.8\textwidth]{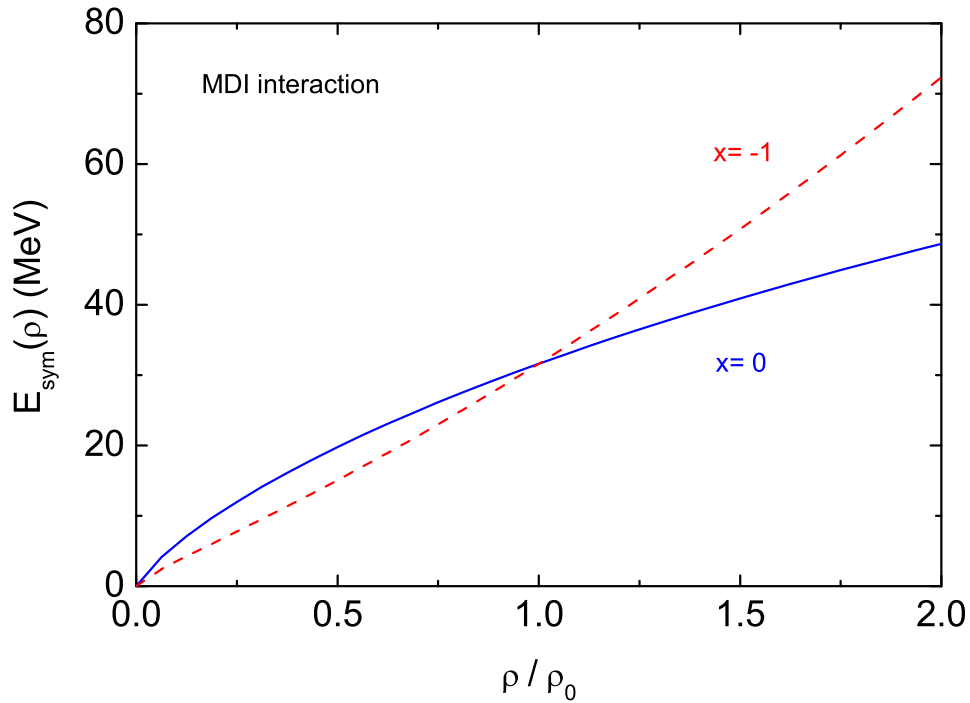}
\end{center}
\caption{{\protect\small (Color online) Density dependence of nuclear
symmetry energy using the MDI interaction with }$x=0${\protect\small \ and }$%
x=-1${\protect\small . Taken from \protect\cite{li05b}.}}
\label{sym}
\end{figure}

According to essentially all microscopic model calculations, see e.g., \cite%
{bom,zuo99}, the EOS for isospin asymmetric nuclear matter can be expressed
as
\begin{equation}
E(\rho ,\delta )=E(\rho ,0)+E_{\text{sym}}(\rho )\delta ^{2}+\mathcal{O}%
(\delta ^{4}),
\end{equation}%
where $E(\rho ,0)$ is the energy per nucleon of symmetric nuclear
matter, and $E_{\text{sym}}(\rho )$ is the nuclear symmetry
energy. With the single particle potential $U(\rho ,\delta
,\mathbf{p},\tau )$, for a given
value $x$, one can readily calculate the symmetry energy $E_{\text{sym}%
}(\rho )$ as a function of density. Noticing that the isospin diffusion data
from NSCL/MSU have constrained the value of $x$ to be between $0$ and $-1$
for nuclear matter densities less than about $1.2\rho _{0}$ \cite%
{chen04,li05}, in the present work we thus consider only the two values of $%
x=0$ and $x=-1$. Shown in Fig.\ \ref{sym} is the density
dependence of the nuclear symmetry energy with the two $x$
parameters. It is seen that the case of $x=0$ gives a softer
symmetry energy than that of $x=-1$ and the difference becomes
larger at higher densities.

\section{Results and discussions}
\begin{figure}[th]
\begin{center}
\includegraphics[width=0.8\textwidth]{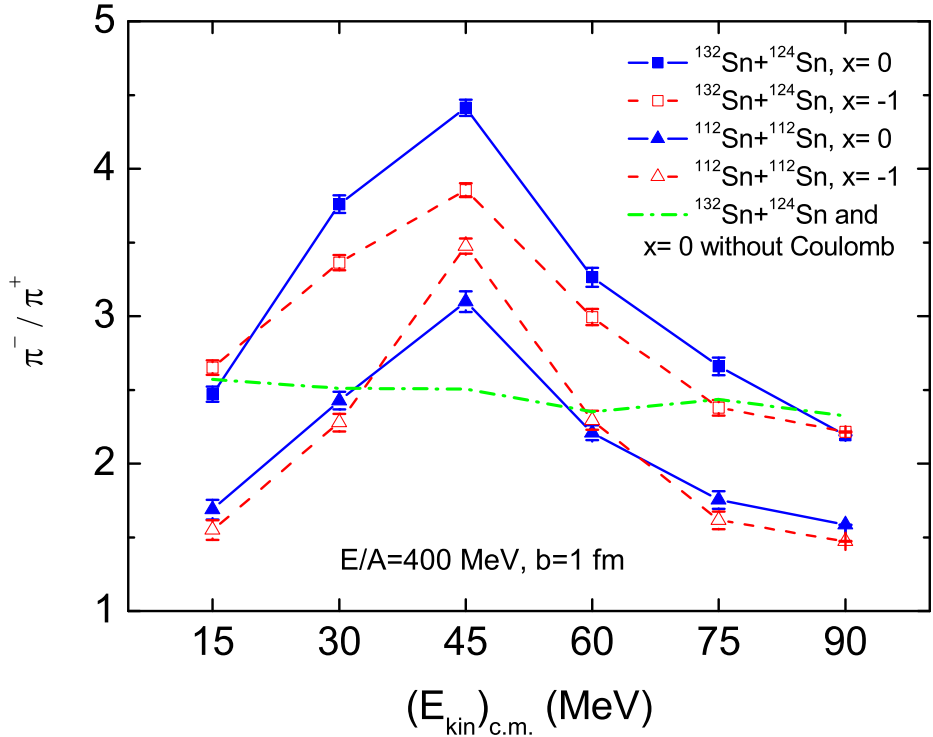}
\end{center}
\caption{{\protect\small (Color online) Kinetic energy distribution of the
single }$\protect\pi ^{-}/\protect\pi ^{+}${\protect\small \ ratio for }$%
^{132}${\protect\small Sn+}$^{124}${\protect\small Sn and }$^{112}$%
{\protect\small Sn+}$^{112}${\protect\small Sn at a beam energy of }$400$%
{\protect\small \ MeV/nucleon and an impact parameter of }$b=1$%
{\protect\small \ fm with the stiff (}$x=-1${\protect\small ) and soft (}$x=0
${\protect\small ) symmetry energies. The dash-dotted line is the single }$%
\protect\pi ^{-}/\protect\pi ^{+}${\protect\small \ ratio obtained by
turning off the Coulomb potentials in the }$^{132}${\protect\small Sn+}$%
^{124}${\protect\small Sn reaction.}}
\label{Rpion}
\end{figure}

It was found earlier that the single $\pi ^{-}/\pi ^{+}$ ratio from
heavy-ion collisions induced by neutron-rich nuclei can be used to probe the
high density behavior of the nuclear symmetry energy \cite%
{ba02a,gai04,qli05a,qli05b}. In Fig.\ \ref{Rpion}, we show the kinetic
energy distribution of the single $\pi ^{-}/\pi ^{+}$ ratio for the
reactions of $^{132}$Sn+$^{124}$Sn and $^{112}$Sn+$^{112}$Sn at a beam
energy of $400$ MeV/nucleon and an impact parameter of $b=1$ fm with the
stiff ($x=-1$) and soft ($x=0$) symmetry energy, respectively. In order to
obtain good statistics, we used $12000$ events for each reaction in the
present work. It is seen that the overall magnitude of $\pi ^{-}/\pi ^{+}$
ratio is larger for the neutron-rich system $^{132}$Sn+$^{124}$Sn than for
the neutron-deficient system $^{112}$Sn+$^{112}$Sn as expected. For the
reaction $^{112}$Sn+$^{112}$Sn the single $\pi ^{-}/\pi ^{+}$ ratio is not
so sensitive to the symmetry energy due to the small isospin asymmetry.
However, for the neutron-rich system $^{132}$Sn+$^{124}$Sn the single $\pi
^{-}/\pi ^{+}$ ratio is sensitive to the symmetry energy and this is
consistent with previous studies \cite{ba02a,gai04,qli05a,qli05b}. It is
further seen that the soft symmetry energy ($x=0$) leads to a larger single $%
\pi ^{-}/\pi ^{+}$ ratio than the stiff one ($x=-1$). This is mainly because
the high density region where most pions are produced are more neutron-rich
with the softer symmetry energy as a result of the isospin fractionation
\cite{ba02a}.

From Fig.\ \ref{Rpion}, it is also interesting to see that the single $\pi
^{-}/\pi ^{+}$ ratio exhibits a peak at a pion kinetic energy of about $45$
MeV in all cases considered here. In order to understand the origin of this
peak, we also calculated the single $\pi ^{-}/\pi ^{+}$ ratios in both
reactions by turning off the Coulomb potentials for all charged particles.
As an example, shown in Fig.\ \ref{Rpion} with the dash-dotted line is the
single $\pi ^{-}/\pi ^{+}$ ratio obtained by turning off the Coulomb
potentials in the $^{132}$Sn+$^{124}$Sn reaction. It is seen that the single
$\pi ^{-}/\pi ^{+}$ ratio now becomes approximately a constant of about $2.4$%
. The latter is what one expects based on the $\Delta $ resonance model \cite%
{stock}. According to the latter the $\pi ^{-}/\pi ^{+}$ ratio is
approximately $(5N^{2}+NZ)/(5Z^{2}+NZ)\approx (N/Z)^{2}$ in central
heavy-ion reactions with $N$ and $Z$ being the total neutron and proton
numbers in the participant region \cite{stock}. For central $^{132}$Sn+$%
^{124}$Sn reactions, the value is about $2.43$. At $400$ MeV/nucleon, pions
are almost exclusively produced via the $\Delta $ resonances \cite{ba91}, it
thus should not be a surprise to see the agreement with the $\Delta $
resonance model expectation. The comparison of calculations with and without
the Coulomb potentials indicates clearly that the peak observed in the
single $\pi ^{-}/\pi ^{+}$ ratio is indeed due to the Coulomb effects. The $%
\pi ^{-}/\pi ^{+}$ ratio carries some information about the
symmetry energy mainly because it is sensitive to the isospin
asymmetry of the nucleonic matter where pions are produced. This
information might be distorted but is not completely lost because
of the Coulomb interactions of pions with other particles. It is
thus natural to look for signals of the symmetry energy in
kinematic regions where the $\pi ^{-}/\pi ^{+}$ ratio reaches its
maximum. In this regard, the Coulomb peak is actually very useful
for studying the symmetry energy. It is necessary to stress that
in many situations the Coulomb peak will simply appear at zero
instead of a finite kinetic energy. One would then need to
concentrate on the $\pi ^{-}/\pi ^{+}$ ratio of low energy pions.
In addition, it should be mentioned that most of pions are
produced in the high density nucleonic matter (about $2\rho _{0}$)
through the $\Delta $ resonances and thus carry important
information about the high density behavior of the symmetry
energy. Since the pions at lower kinetic energies around the
Coulomb peak experience many rescatterings with nucleons at high
or low densities and the charged pions considered here also feel
the Coulomb potential from protons at different densities, the
information on high density symmetry energy from the lower energy
pions may be distorted partially by the low density behavior of
the symmetry energy \cite{qli05a}. However, in the present work,
the soft ($x=0$) and stiff ($x=-1$) symmetry energies have slight
difference at low densities and the large difference appears at
high densities (about $2\rho _{0}$) as shown in Fig.\ \ref{sym}
and we thus expect the observed symmetry energy effects on the
energy dependence of the $\pi ^{-}/\pi ^{+}$ ratio mainly reflect
(though not completely) information on the high density behavior
of the symmetry energy.
\begin{figure}[th]
\begin{center}
\includegraphics[width=0.8\textwidth]{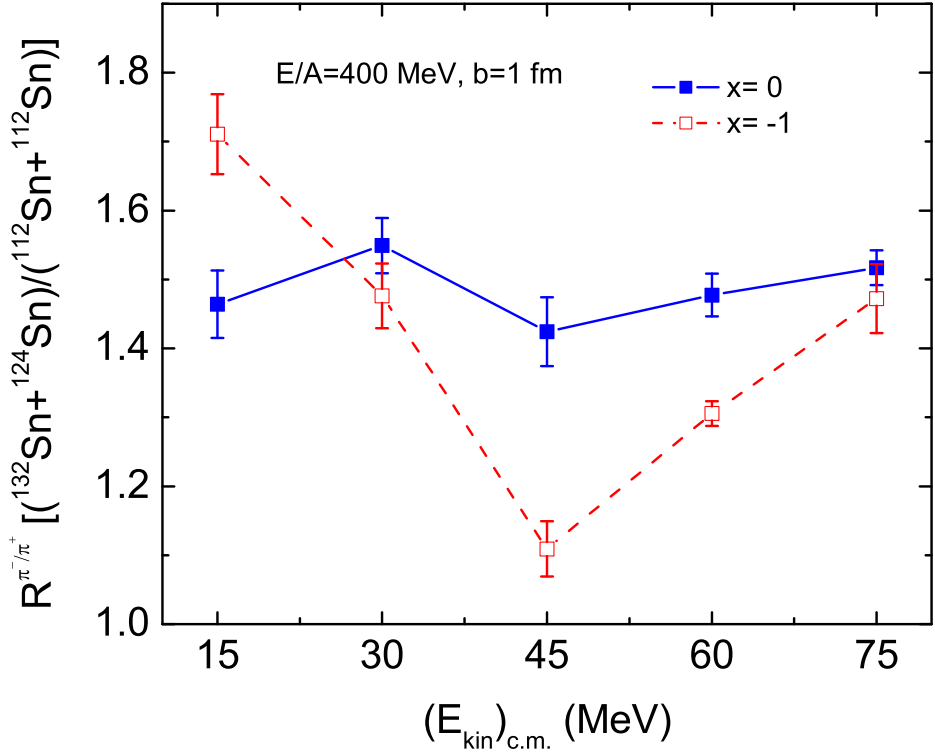}
\end{center}
\caption{{\protect\small (Color online) Kinetic energy dependence of the
double }$\protect\pi ^{-}/\protect\pi ^{+}${\protect\small \ ratio of }$%
^{132}${\protect\small Sn+}$^{124}${\protect\small Sn over }$^{112}$%
{\protect\small Sn+}$^{112}${\protect\small Sn at a beam energy of }$400$%
{\protect\small \ MeV/nucleon and an impact parameter }$b=1${\protect\small %
\ fm with the stiff (}$x=-1${\protect\small ) and soft (}$x=0$%
{\protect\small ) symmetry energies.}}
\label{DRpion}
\end{figure}

To reduce the systematic errors, it is more useful to study the
double $\pi ^{-}/\pi ^{+}$ ratio in the reactions of
$^{132}$Sn+$^{124}$Sn and $^{112}$Sn+$^{112}$Sn. A practically
critical question is whether the sensitivity to the symmetry
energy observed in the single $\pi ^{-}/\pi ^{+}$ ratio can be
sustained by the double ratio. To answer this question we examine
in Fig.\ \ref{DRpion} the double $\pi ^{-}/\pi ^{+}$ ratio for the
two reactions considered. It is seen that the kinetic energy
dependence of the double $\pi ^{-}/\pi ^{+}$ ratio is rather
different for the stiff ($x=-1 $) and soft ($x=0$) symmetry
energies. The double $\pi ^{-}/\pi ^{+}$ ratio is quite flat for
$x=0$ while displaying a concave structure for $x=-1$ around the
Coulomb peak. These different behaviors can be understood from the
corresponding single $\pi ^{-}/\pi ^{+}$ ratios in the two
reactions as shown in Fig. \ref{Rpion}. It is reassuring to see
that around the Coulomb peak the double $\pi ^{-}/\pi ^{+}$ ratio
is still sensitive to the symmetry energy. Compared with the
single $\pi ^{-}/\pi ^{+}$ ratio, the kinetic energy dependence of
the double $\pi ^{-}/\pi ^{+}$ ratio becomes weaker.
This is because the effects of the Coulomb potentials are reduced in the double $%
\pi ^{-}/\pi ^{+}$ ratio. We note that the double $\pi ^{-}/\pi
^{+}$ ratio displays an opposite symmetry energy dependence
compared with the double $n/p$ ratio for free nucleons shown in
Ref. \cite{li05b}. This is understandable since the soft symmetry
energy leads to a neutron-richer dense matter in the
heavy-ion collisions induced by neutron-rich nuclei and thus more $\pi ^{-}$%
's would be produced due to more neutron-neutron inelastic scatterings.
Since the soft symmetry energy leads to a neutron-richer dense nucleonic
matter, the $n/p$ ratio for free nucleons is therefore expected to be smaller
due to the charge conservation.

\section{Summary}
In summary, using the isospin- and momentum-dependent hadronic transport
model IBUU04, we studied the single and double $\pi ^{-}/\pi ^{+}$ ratios
and their dependence on the nuclear symmetry energy in the central reactions
of $^{132}$Sn+$^{124}$Sn and $^{112}$Sn+$^{112}$Sn at a beam energy of $400$
MeV/nucleon. We found that the double $\pi ^{-}/\pi ^{+}$ ratio retains the
same sensitivity to the symmetry energy as shown by the single $\pi ^{-}/\pi
^{+}$ ratio around its Coulomb peak in the neutron-richer system involved.
These features are useful for extracting information about the nuclear
symmetry energy from future experimental data. Because the double $\pi
^{-}/\pi ^{+}$ ratio can reduce significantly the systematic errors, it is
thus a more effective probe for the high-density behavior of the nuclear
symmetry energy.

\section*{Acknowledgments}
The work is supported by the US National Science Foundation under Grant No.
PHY-0354572, PHY0456890 and the NASA-Arkansas Space Grants Consortium Award
ASU15154, the National Natural Science Foundation of China under Grant No.
10105008, 10575071, 10575119, 10235030, the Chinese Academy of Science
Knowledge Innovation Project (KJCX2-SW-N02), Major State Basic Research
Development Program (G2000077400), and the Chinese Ministry of Science and
Technology.

\end{document}